\documentclass{elsart}
\usepackage{natbib}
\usepackage{epsfig} 
\begin{document}
\runauthor{Krawczynski}
\begin{frontmatter}
\title{TeV Blazars --- Observations and Models\vspace*{-1.2cm}}
\author[washu]{Henric Krawczynski}
\address[washu]{Washington University in St. Louis, Physics Department, 
1 Brookings Drive CB 1105,St. Louis, MO 63130\vspace*{-3.6ex}}
\begin{abstract}
Since the first TeV blazar Markarian (Mrk) 421 was detected in 1992, 
the number of established TeV $\gamma$-ray emitting BL Lac objects has 
grown to 6, with redshifts ranging from 0.031 (Mrk 421) to 0.129 
(H 1426+428). The intensive study of these sources has had a major 
impact on our understanding of the blazar phenomenon.
The most notable observational results have been extremely fast
large amplitude flux and spectral variability on hour time scales, 
and a pronounced X-ray - TeV $\gamma$-ray flux correlation.
In this paper we discuss recent observational results and report
on progress in their theoretical interpretation.
\vspace*{-0.4cm}
\end{abstract}
\end{frontmatter}
\vspace*{-1.3cm}
\section{Introduction -- TeV Blazars}
\vspace*{-0.9cm}
The EGRET {\it (Energetic Gamma Ray Experiment Telescope)} 
detector on board of the {\it Compton Gamma-Ray Observatory} 
discovered strong MeV $\gamma$-ray emission from 66 blazars, 
mainly from Flat Spectrum Radio Quasars and Unidentified Flat Spectrum Radio Sources  \cite{Hart:99}.
Ground-based Cherenkov telescopes discovered TeV $\gamma$-ray emission from
6 blazars, 4 of which have not been detected by EGRET.
All the established TeV detections belong to the class of BL Lac objects, 
blazars with relatively low luminosity but with Spectral Energy Distributions 
(SEDs) that peak at extremely high energies (see Table \ref{detect}).\\[.5ex]
\begin{table}[bt]
\begin{center}
\begin{tabular}{|c|c|c|} \hline
Source & $z$ & Discovery \& Confirmation\\ \hline
 Mrk 421     & 0.031  & Punch et al.\     1992, Petry et al.\     1996  \\[-1.4ex]
 Mrk 501     & 0.034  & Quinn et al.\     1996, Bradbury et al.\  1997  \\[-1.4ex]
 1ES 2344+514    & 0.044  & Catanese et al.\  1998, Tluczykont et al.\ 2003  \\[-1.4ex]
 1ES 1959+650    & 0.047  & {\scriptsize Nishiyama\,et\,al.\,1999,\,Holder\,et\,al.\,2003,\,Aharonian\,et\,al.\,2003}\\[-1.4ex]
 PKS 2155-304    & 0.116  & Chadwick et al.\  1999, Hinton et al.\    2003  \\[-1.4ex]
 H 1426+428    & 0.129  & Horan et al.\     2002, Aharonian et al.\ 2002a \\ \hline
\end{tabular}
\end{center}
\caption{TeV blazars with a significant detection by at least 2 experiments (as of July 2003).}
\label{detect}
\end{table}
The large detection area of Cherenkov telescopes of 
$\sim$10$^5$ m$^2$ makes it possible to sample the $\gamma$-ray lightcurves
with a time resolution of several minutes.
Large amplitude flux variability on 30 min time scales 
implies that the TeV emission originates from a small region very near the 
supermassive black hole \cite{Gaid:96,Kraw:01a}. 
The relativistic Doppler Factor of the emitting plasma is defined as 
$\delta_{\rm j}^{-1}\,=\,\Gamma(1-\beta\,\cos{(\theta)})$, 
where $\Gamma$ is the bulk Lorentz factor of the emitting plasma, 
$\beta$ is its bulk velocity in units of the speed of light, 
and $\theta$ is the angle between jet axis and the line of sight as 
measured in the observer frame.
The requirement that the emitting volume is optically thin for the 
TeV $\gamma$-radiation, together with assumptions about the optical to UV
energy spectrum of co-spatially emitted synchrotron emission gives a lower limit
of $\delta_{\rm j}\,>$ 9 for the emitting plasma \cite{Gaid:96}.
The Lorentz boost of the observed jet luminosity by the factor 
$\delta_{\rm j}^{\,\,4}$ makes the rather weak jet emission readily 
detectable.\\[0.5ex]
The main scientific drivers for broadband observations of TeV blazars
are: (i) the study of the matter, energy content, and structure
of jets from Active Galactic Nuclei (AGNs), (ii) the 
time resolved investigation of particle acceleration processes, 
(iii) the research on the connection between the accreting supermassive 
black hole system and the jet parameters, and 
(iv) the measurement of the total infrared/optical luminosity of the universe.\\[0.5ex]
In this paper we give an update on the status of the observations
(Sect.\ \ref{obs}) and the four major 
scientific topics (Sects.\ \ref{t1}--\ref{t4}).
We conclude with an outlook in Sect.\ \ref{ol}.
We will focus on  observations and modeling of TeV blazars
(blazars detected by TeV instruments).
Reviews on observations and models of MeV/GeV blazars
(detected by EGRET) have been given in \cite{Siko:01b,Copp:99};
recent overviews of TeV astronomy have been given in \cite{Ong:03,Buck:02}.
\vspace*{-1.2cm}
\section{Observations}
\label{obs}
\begin{figure}[t]
\begin{minipage}{9.3cm}
\epsfig{file=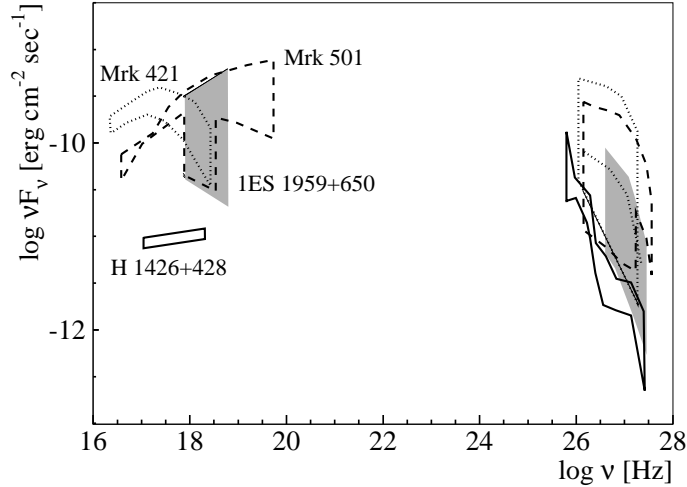,width=9.0cm}
\end{minipage}
\begin{minipage}{4.3cm}
\vspace*{-0.2cm}
\caption{\label{seds} Simultaneous and non-simultaneous X-ray and TeV $\gamma$-ray energy spectra of the
4 TeV blazars with measured TeV $\gamma$-ray energy spectra. The regions show the range of values 
that have been observed with BeppoSAX, RXTE and Cherenkov Telescopes
(from \cite{Kraw:03}).
}
\end{minipage}
\vspace{-0.15cm}
\end{figure}

\vspace*{-0.9cm}
TeV $\gamma$-ray observations have achieved a state of maturity with 
good agreement in fluxes and reasonable agreement in energy spectra between
different experiments.
The measurements of the steady flux from the Crab Nebula with the CAT, HEGRA, and Whipple 
experiments lie within 18.5\% from the common mean comparable to the
systematic error of about 22\% quoted by the groups \cite{Piro:01,Ahar:00,Hill:98}. 
The spectral indices agree to within 0.2 from the common mean, a difference that 
is larger than the systematic errors of between 0.04 and 0.06 estimated by the 3 groups.
Systematic error on flux and spectral changes from one day to the next
are estimated to be about 10\% and 0.05, respectively. \\[.5ex]
High-quality TeV $\gamma$-ray energy spectra have now been measured for 
Mrk 421, Mrk 501, H~1426+428, and 1ES~1959+650. Fig.\ \ref{seds} compares the X-ray 
and TeV $\gamma$-ray SEDs of these 4 sources.
Compared to the other 3 blazars, the low-energy (synchrotron) component of Mrk 421 
peaks at lower frequencies. Comparison of the high-energy TeV $\gamma$-ray energy spectra is not straight forward as it depends on the
uncertain extent of extragalactic absorption (see Sect.\ \ref{t4}).
Spectral variability has been established for the two strongest ones, 
Mrk 421 \cite{Kren:02,Ahar:02b} and Mrk 501 \cite{Djan:99,Ahar:01}.
The significance of measurements of spectral changes can not be 
overestimated. Spectral changes (like flux changes) are free of the still
substantially uncertain extent of extragalactic $\gamma$-ray absorption. 
Modeling of simultaneously taken X-ray and $\gamma$-ray data with well
measured spectral changes in both bands has not yet been performed
but has the potential of breaking basically all model degeneracies.\\[.5ex]
Several extensive multiwavelength campaigns have been performed.
The correlation of X-ray and TeV $\gamma$-ray fluxes had first been
seen in Mrk 421 data \cite{Buck:96,Taka:96}.
In the meantime, correlations with high statistical significances have been observed 
from Mrk 501 \cite{Kraw:00,Samb:00} and from Mrk 421 \cite{Taka:00,Foss:03}.
A more recent campaign on the TeV blazar 1ES~1959+650 shows an 
orphan $\gamma$-ray flare without an X-ray or optical counterpart  \cite{Kraw:03}. 
The orphan flare shows that the simplest models (1-zone Synchrotron Self-Compton
models with a randomly oriented magnetic fields and isotropic electron distributions) 
are not able to account for the data. \\[.5ex]
Simultaneous measurements of blazar energy spectra with imaging Cherenkov telescopes 
and solar array Cherenkov detectors have the potential of producing highly interesting 
results. First lightcurves from STACEE \cite{Boon:02} and a first energy spectrum from 
CELESTE \cite{Piro:03} have been reported recently.
\vspace*{-1.2cm}
\section{Emission Models and Implications for the Jet Structure}
\label{t1}
\vspace*{-0.9cm}
{\bf Synchrotron-Compton Models:}
In Synchrotron Compton (SC) models of TeV blazars the radio to X-ray emission is produced as synchrotron radiation 
from a population of non-thermal electrons (and possibly positrons).
The same electron population emits $\gamma$-rays through Inverse Compton processes  
by electrons scattering synchrotron photons in so-called Synchrotron Self-Compton (SSC) models
or with ``external'' photons that originate outside the jet in  so-called External Compton (EC) models
(see \cite{Siko:01b} and references therein).
In the case of BL Lac objects, the lack of strong emission lines is commonly taken as
evidence that ambient photon fields are not important, and that SSC models rather than 
EC models are more likely to explain the data.\\[0.5ex]
A method to distinguish between SSC and EC models is to measure the time lag $\Delta t_{\rm S-IC}$
between X-ray and $\gamma$-ray flux variability \cite{Copp:99b}.
In SSC models one expects a time-lag of the order of $R$ $c^{-1}$ 
$\delta_{\rm j}^{-1}$
with $R$ the radius of the emission volume and $c$ the speed of light.
The time lag originates from the fact that a population of 
freshly accelerated electrons immediately radiates synchrotron emission; 
in contrast, the Inverse Compton component peaks only after the new synchrotron 
seed photons had time to propagate through the source.
No such time lag has yet been observed with high statistical confidence.\\[0.5ex]
TeV blazar data have been modeled intensively with a variety of codes. 
The more simpler ones assume 1-zone homogeneous emission volumes and use phenomenological 
electron energy spectra to fit the broadband data (e.g.\ \cite{Inou:96,Taka:00,Tave:01}).
Modeling with time-dependent  self-consistently evolved codes (e.g.\ \cite{Copp:92,Mast:97,Kraw:02,Boet:02}) 
assures that the energy spectra are {physically 
realizable} from initial acceleration energy spectra.
Time dependent modeling has to be used whenever the flare duration 
is comparable or shorter than the longest time scale of the microprocesses 
that modify the electron energy spectra (e.g.\ acceleration energy gains, 
radiative energy gains and losses).\\[0.5ex]
Several conclusions can be drawn from the SSC and EC modeling performed by various authors:
(i) relatively simple 1-component and 2-component SSC models are able to explain most multiwavelength data; 
(ii) even for the nearest blazars with $z\,\approx$ 0.03,
extragalactic absorption has to be taken into account and substantially
modifies the regions in the parameter space that correspond to acceptable solutions 
(e.g.\ \cite{Ahar:99a,Bedn:99});
(iii) the data prefer high jet Doppler factor of between 50 and 100
\cite{Mast:97,Kraw:01a};
(iv) in the jet frame the energy density of non-thermal particles is much higher than the
energy density of the magnetic field (the jet plasma is a low $\sigma$-plasma) 
(e.g.\ \cite{Kino:02,Kraw:02}); 
(v) only a small fraction ($< 10^{-2}$) of the energy transported by the jet is converted 
into radiation (e.g.\ \cite{Kino:02,Kraw:02}).\\[0.5ex]
The last two conclusions constrain the structure of the jet at its base.
While electromagnetic models of jet formation produce high-$\sigma$ 
jets with large electromagnetic field 
energy to particle energy densities, SSC models indicate low-$\sigma$ jets. 
The low radiative efficiency indicates that the sub-parsec jet can smoothly evolve into the parsec-scale jet, 
without the need for plasma re-collimation or re-acceleration that would have to follow a more efficient
conversion of bulk plasma energy into random particle energy and radiation (see also \cite{Celo:93}).\\[0.5ex]
{\bf Hadronic Models:}
In hadronic models the continuum emission is explained by hadronic interactions of a highly relativistic 
baryonic outflow which sweeps up ambient matter \cite{Pohl:00}, by interactions of high-energy protons 
with gas clouds moving across the jet \cite{Dar:97}, or by interactions of Ultra High Energy 
protons with ambient photons \cite{Mann:93}, with the jet magnetic field \cite{Ahar:00a}, 
or with both \cite{Muec:02}. Although attractive since they involve Ultra High Energy Cosmic Ray 
acceleration, hadronic models have difficulties to explain the good X-ray/TeV $\gamma$-ray flux correlations 
found for Mrk 421 and Mrk 501. Furthermore, little work has yet been done to see if proton models can 
explain more than single snapshots of data. 
\vspace*{-1.2cm}
\section{Particle Acceleration}
\label{t2}
\vspace*{-0.9cm}
Among the many particle acceleration mechanisms that are discussed, Fermi acceleration
at strong Magnetohydrodynamic (MHD) shocks has received most attention. Indeed, the data are largely consistent
with an electron acceleration energy spectrum $dN/dE\,\propto$ $E^{-p}$ with $p$ between 2 and 2.23, 
expected for non-relativistic and ultra-relativistic shocks, respectively (see \cite{Kirk:99a} and references therein).
Based on a time-dependent SC code Kirk \& Mastichiadis (1999) have shown that Fermi energy 
gains and radiative energy losses can result in characteristic hardness--intensity correlations 
during individual flares \cite{Kirk:99}.
X-ray and recently also TeV $\gamma$-ray data do show the predicted signatures:
during flares the sources seem to go through clockwise and anti-clockwise loops in the X-ray or 
$\gamma$-ray hardness--intensity planes (e.g.\ \cite{Taka:96,Horn:02}).
However, the same source shows a wide range of different signatures (e.g.\ \cite{Taka:00}),
indicating that the relative length of the characteristic time scales of particle acceleration and 
radiative cooling change from flare to flare. 
The lack of a prevailing signature has cast doubts on whether ``cooling loops'' and 
``particle acceleration loops''  have really been observed.\\[0.5ex]
Most SC models of TeV blazars invoke a minimum Lorentz factor $\gamma_{\rm min}$ 
of accelerated particles on the order of $10^5$ to account for the data (e.g.\ \cite{Kraw:02} 
and references therein).
Such high $\gamma_{\rm min}$-values could either result from a yet unknown
pre-acceleration mechanism, or, from an upstream bulk Lorentz factor on the 
order of $10^5$. A low-entropy, extremely relativistic upstream plasma could be detectable by its
Inverse Compton emission if exposed to a suitable external radiation field.
\vspace*{-1.2cm}
\section{Black Hole - Jet Connection}
\label{t3}
\vspace*{-1cm}
\begin{figure}[t]
\begin{center}
\epsfig{file=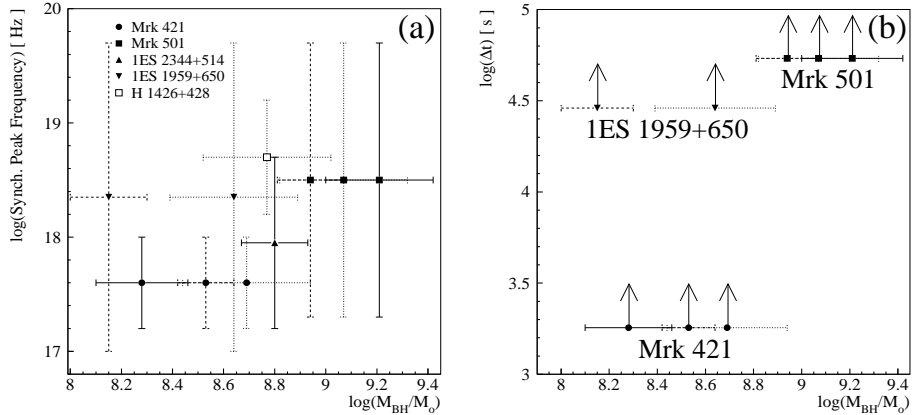,width=12cm}
\end{center}
\vspace*{-0.4cm}
\caption{\label{bm} \small Correlation of estimated black hole mass and the peak of the Synchrotron SED
(a) and observed TeV $\gamma$-ray flux variability time-scales (b) for the 6 well established 
TeV blazars. Black hole masses from stellar velocity dispersion 
are shown by solid (from \cite{Bart:03}) and dashed (from \cite{Falo:02}) error bars.
Black hole masses from the galactic bulge luminosity (from \cite{Falo:02}) are shown by dotted error bars.
Vertical error bars show the range of observed values. 
Horizonthal error bars are statistical errors in the case of solid and dashed lines, for the dotted estimates no
statistical errors have ben published and we assumed $\Delta log(M_\bullet)\,=$ 0.25 (figures from \cite{Kraw:03}).}\vspace*{-0.05cm}
\end{figure}
From the theoretical side, the connection between black hole and jet properties is far from being
understood. Purely electromagnetic \cite{Blan:76,Love:76} and GR-electromagnetic models \cite{Blan:77} 
of jet formation face the well known $\sigma$-problem, namely that they predict Poynting flux 
dominated energy transport while SC models indicate strongly particle energy dominated jets.
Hydrodynamic, MHD, and GR-MHD scenarios of jet formation 
(e.g.\ \cite{Bege:84,Koid:99}) face the difficulty that they predict jets with bulk Lorentz 
factors $\Gamma<10$ while SC models favor $\Gamma\gg 10$.\\[0.5ex]
From the observational side, first studies of the black hole--jet connection are just
becoming possible based on high-quality black hole mass estimates from galactic stellar 
velocity dispersion measurements \cite{Falo:02,Bart:03}.
The studies performed so far do not indicate any correlation of the black hole mass and
the peak location or peak luminosity of the low-energy or high-energy SEDs,
the time scale of flux variability (see Fig.\ \ref{bm}), or the flare duty cycle.
The correlations may be masked by variation of parameters like jet viewing angle, jet magnetic field, 
or intensity of the ambient photon field \cite{Bart:03,Kraw:03}.
Alternatively, the black hole mass estimators may not work in the case of blazars, 
as discussed in \cite{Bart:03}.\vspace*{-1.3cm}
\section{Measurement of the Cosmic Infrared Background}
\label{t4}
\vspace*{-0.9cm}
Extragalactic absorption of TeV $\gamma$-rays by the Cosmic Infrared Background (CIB) 
and the Cosmic Optical Background (COB) in $\gamma_{\rm TeV}+\gamma_{\rm CIB/COB}\,\rightarrow$ $e^+\,e^-$
pairproduction processes \cite{Niki:62,Goul:65,Stec:92,Prim:01,Jage:02}
substantially complicates the study of the astrophysics of blazar jets.
However, the absorption effect allows us to use TeV $\gamma$-ray 
observations to measure the energy spectrum of the CIB and the COB.
These backgrounds are extremely interesting owing to their close relation to the total electromagnetic 
luminosity of the universe since decoupling of matter and radiation 
$\sim$300,000 years after the big bang. \\[0.5ex]
The TeV energy spectra of extragalactic sources carry the imprint of extragalactic absorption
in the form of ``characteristic high-energy cutoffs'' in the 30 GeV to 10 TeV energy range.
The measurements of these cutoffs makes it possible to constrain the energy spectrum of the 
infrared to optical background radiation. These constraints are very valuable, 
as direct measurements are plagued by large systematic errors, owing to strong foreground 
emission from our solar system and the Milky Way (e.g.\ \cite{Haus:01}).
While the spectral measurement of the nearby blazars Mrk 421 and Mrk 501 ($z\,\approx$ 0.03) 
have been used to set upper limits on the CIB intensity in the 1-20~$\mu$m wavelength range 
\cite{Bill:98}, the recent measurement of the energy spectrum from the more distant
blazar H 1426+428 ($z\,=$ 0.129) has given the first demonstration of the full potential 
of TeV $\gamma$-ray observations to decide between different background models \cite{Ahar:03a}.\\[0.5ex]
Further progress may best be accomplished with a two-pronged approach:
(i) Detailed source modeling makes it possible to disentangle, at least partially, source 
inherent high-energy cutoffs from those caused by extragalactic extinction and thus to 
improve on the constraints from each single source \cite{Copp:99,Kraw:02};
(ii) since the spectral cutoffs from extragalactic absorption are expected to occur at energies which depend only 
on source redshift, a statistical analysis of the constraints from many sources 
can be used to test and strengthen the constraints from individual sources. 
\vspace*{-1.2cm}
\section{Outlook}
\vspace*{-0.9cm}
\label{ol}
Over the last several years, the field of TeV blazar studies has benefited greatly from 
detections of new sources and from very intensive multiwavelength campaigns.
Further progress may derive from extending the time scales of the broadband observations.
High-accuracy measurement of Gamma-Ray Burst (GRB) afterglows is boosting the theoretical progress 
in the case of GRBs. In the same way, detection of increased flux levels in the radio to UV bands following
major blazar flaring phases may prove crucial for pinning down the jet properties.
For this purpose regular broadband monitoring is required over time scales longer than 
the typical 10 days of current multiwavelength campaigns.
As the quality of the data improves, it now becomes apparent that simple one-zone
SC model fit snapshots of the data, but are not able to give us deep insights into the 
jet structure and the nature of the particle acceleration processes.
Most present theoretical work focuses on the emission mechanism and largely neglects the flare 
origin and the development of the plasma that supports the non-thermal particles during the flare.
A notable exception is the recent work concerning the internal shock model \cite{Spad:01,Tani:03}.
In future work, combining 3-dimensional MHD simulations of flares with kinetic simulations of 
the non-thermal particle population would make it possible to explore more complex observational 
signatures than possible with present codes.\\[0.5ex]
In view of the next-generation Cherenkov telescopes CANGAROO III, H.E.S.S., MAGIC, and VERITAS
presently under construction, the upcoming launches of the X-ray observatories SWIFT and ASTRO-E2, 
and the $\gamma$-ray observatories AGILE and GLAST to be launched in 2005 and 2006, respectively,
we can expect a wealth of exciting blazar results in the next 5 years.
\\[0.5ex]
{\it Acknowledgements:} The author gratefully acknowledges support by NASA through the
grant NASA NAG5-12974.\\[-10ex]

\end{document}